\documentclass[twocolumn,twoside]{revtex4}
\usepackage{graphicx}
\usepackage{fancyhdr}
\begin{document}

\title{Future Accelerator-Based Neutrino Beams}

%

\author{Vladimir Shiltsev}
\affiliation{Fermilab, MS312, PO Box 500, Batavia, IL, 60510, USA }

\begin{abstract}
High energy and high beam power accelerators are extensively used for the neutrino physics research. At present, the leading operational facilities are the Fermilab Main Injector complex that delivers over 0.75 MW of 120 GeV protons on the neutrino target, and the J-PARC facility in Japan which recently approached 0.5 MW of the 30 GeV proton beam power. Here we present the status and planned upgrades of the Fermilab and J-PARC accelerators and proposals for the next generation accelerator-based facilities, their challenges and required and ongoing accelerator R\&D programs aimed to address corresponding cost and performance risks. 
\end{abstract}

\maketitle

\thispagestyle{fancy}


\section{FERMILAB AND J-PARC FACILITIES}
Intense high energy particle beams are widely used to  uncover the elusive properties of neutrinos and observe rare processes that probe physics beyond the Standard Model \cite{IF2012, PDG2018}. In particular, the neutrino beams from high-energy proton accelerators, derived from the decays of charged $\pi$ and $K$ mesons, which in turn are created from proton beams striking thick nuclear targets,  have been instrumental discovery tools in particle physics \cite{BK2003, EF2006, SK2007}. Currently, the most powerful accelerators for the neutrino research are under operation at Fermilab (Batavia, IL, USA) and at J-PARC (Tokai, Japan). 

\begin{figure}[h]
\centering
\includegraphics[width=80mm]{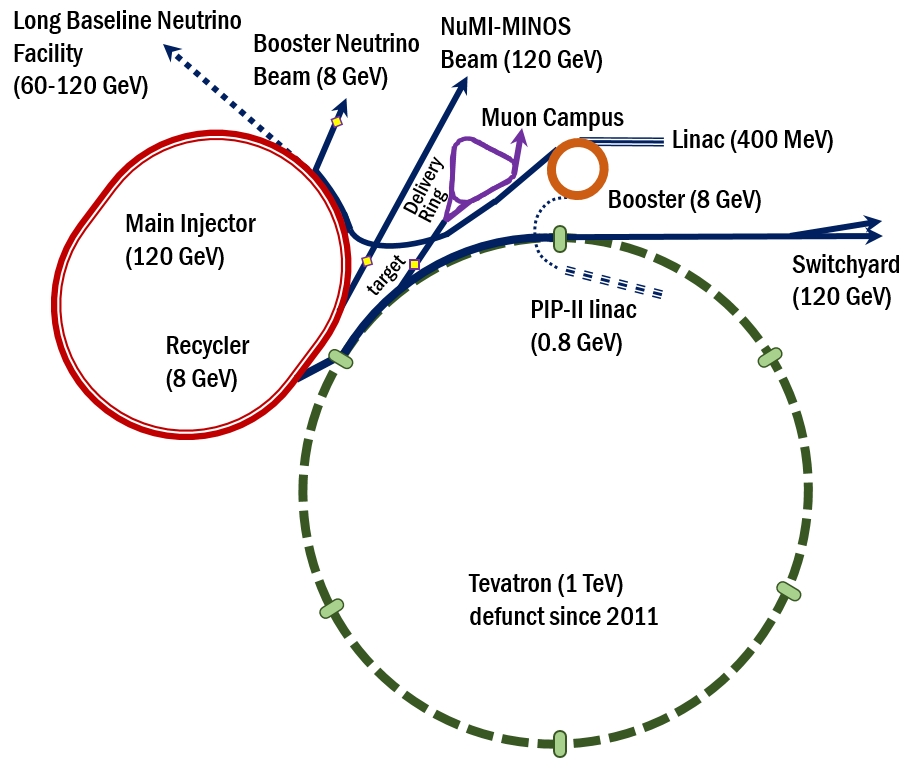}
\caption{Fermilab accelerator complex (from \cite{VS2017}).} 
\label{Fermilab}
\end{figure}
Fermilab accelerator complex - see Fig.\ref{Fermilab} - includes 400 MeV H- pulsed normal-conducting RF linac, 8 GeV proton Booster synchrotron, 8 GeV Recycler storage ring that shares a 3.3 km tunnel with the 120 GeV proton Main Injector (MI) synchrotron, and a 3.1 GeV muon Delivery Ring. A number of beamlines connect the accelerators, bring the beams to fixed targets and to various high energy physics experiments. The most notable future additions (dotted lines) include the LBNF beam line  for DUNE and 0.8 GeV CW-capable SRF PIP-II linac located inside the Tevatron ring and the corresponding beamline for injection into the Booster \cite{VS2017, MC2018}. The complex supports a number of experiments – e.g., the 400 MeV Linac beam is sent to the Mucool Test Area, 8 GeV protons from the Booster are supplied to the 8 GeV Booster Neutrino Beam (BNB), ANNIE, MicroBooNE, MiniBooNE, MITPC, ICARUS (near future), and SBND (future), and to the “muon g-2” and “Mu2e” (near future) muon experiments. The 120 GeV MI proton beam supports neutrino experiments at NuMI (MINERvA, NOvA) and DUNE in the future, as well as the fixed target experiments SeaQuest and test beam facility. See Refs.\cite{FNALIF, EPPSU167} for detailed information on these  experiments. The Fermilab accelerator complex has delivered neutrino beam with over 85\% uptime on average over the past 5 years \cite{MC2018}. 

Construction of the Japan Proton Accelerator Research Complex (J-PARC) was started at the Tokai site of JAEA in 2001 and finished in 2009 \cite{JPARC1}. The facility consists of three accelerators: 400 MeV linac, 350m circumference 3 GeV proton rapid cycling synchrotron (RCS) and a 1.6 km circumference 30 GeV main ring (MR) synchrotron. The design beam power goal for the RCS and MR are 1 MW. The particle research facilities constructed at J-PARC are the Materials and Life Science Facility (MLF) for usage of muon and neutron beams, and the Nuclear and Particle Physics Facility (Hadron Facility) for K-meson (Kaon) beam and the  Neutrino Facility for neutrino beam which is now sent 295 km away to the Super-Kamiokande experiment.
Fig.\ref{FNALJPARC} presents to date progress of these high-energy high- power accelerators for neutrino research: the J-PARC facility in Japan has achieved 475 kW of the 30 GeV proton beam power \cite{JPARC2019}, the Fermilab Main Injector delivers record high 760 kW of 120 GeV protons 
\cite{MC2018}. 

\section{Ways to Increase Beam Power}
Average proton beam power on the neutrino target is scales with the beam energy $E_b$, number of particle per pulse (PPP) $N_{ppp}$ and cycle time $T_{cycle}$ as :
\begin{equation}
P_b = \frac{E_b N_{ppp}}{T_{cycle}}.
\label{pbeam}
\end{equation}
Correspondingly, there are several ways to increase $P_b$:
a) the "brute force" approach - to increase the beam energy, that would require either better or new magnets and RF acceleration system, and/or to decrease the cycle time (again, new magnets and RF might be needed); the key challenges along that path will be the cost of the upgrade (for the reference, the total project cost of the J-PARC facility is about \$1.7B) and significant increase of the facility AC power consumption; b) another approach is to increase the number of protons per pulse (PPP) - where the key challenges mostly associated with numerous beam dynamics issues and, sometimes, with the cost. In both cases one would need reliable horns and targets to shape the secondary beams. There the challenge is that the operational lifetime of the targets gets worse (shorter) with the power 
increase and frequent replacements of them may pose serious impediment on  facility's uptime. 
\begin{figure}[h]
\centering
\includegraphics[width=80mm]{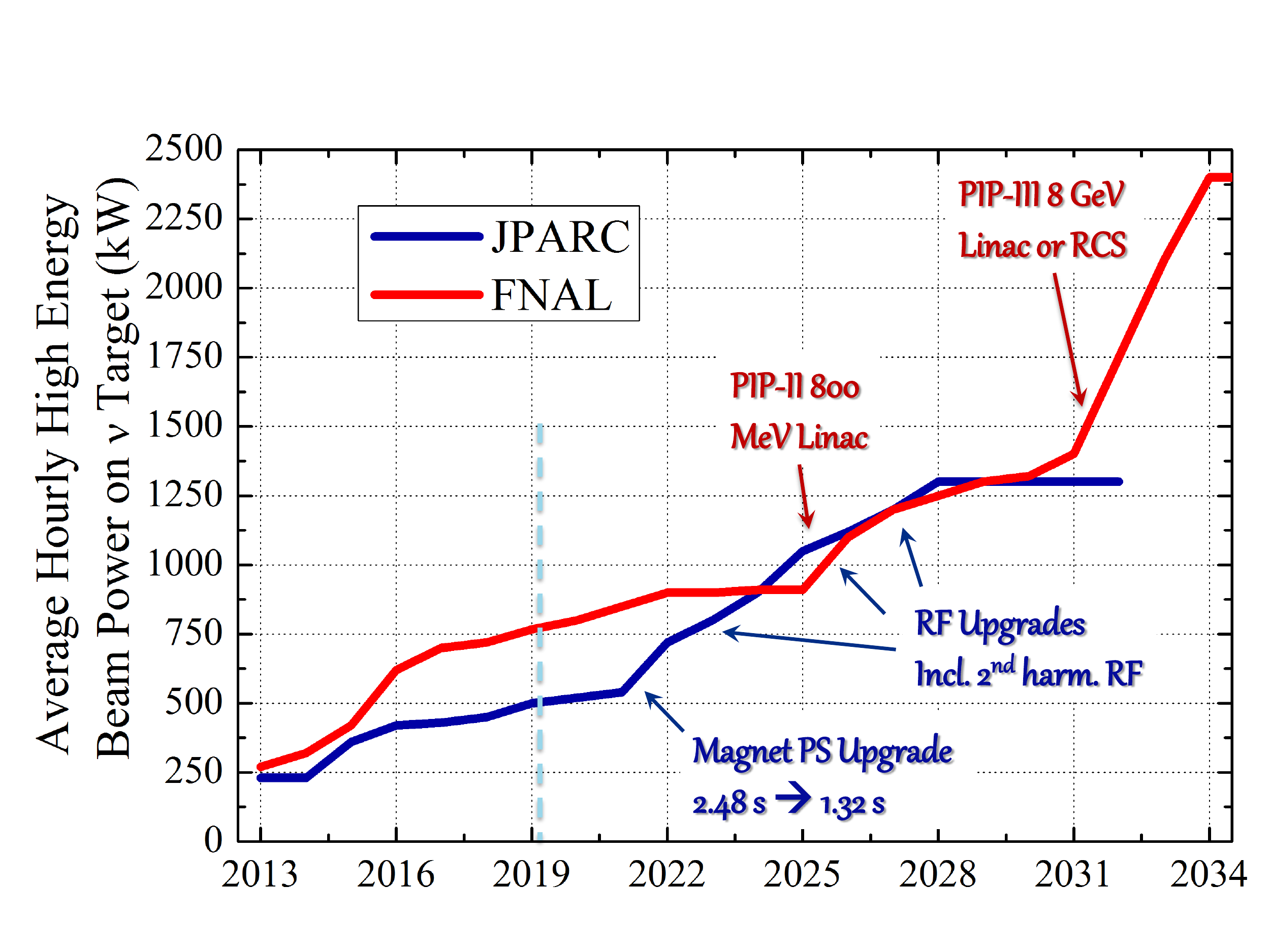}
\caption{Fermilab and J-PARC proton beam power history and upgrade plans.} \label{FNALJPARC}
\end{figure}

Modern high power proton RCSs are quite expensive, their cost is second only to that of colliders and the magnets and associated power supplies constitute 40\% to 50 \% of the cost while the tunable frequency RF cavities and RF power sources are another 15-25\%. Also, the magnets and RF cavities are major consumers of the AC power: e.g. FNAL MI 1.7 T 3 T/s magnets use 9 MW of power, while the RF system uses another 2.5 MW. The RCS overall power efficiency is a serious issue - e.g. production of some 0.5 MW beams at J-PARC comes with about 40 MW site AC power. Still, the upgrade of the J-PARC main magnet power supply will allow reduction of the cycle time from 2.48 s to 1.32 s and proportional increase in $P_b$ - see Fig.\ref{FNALJPARC}. 

There are several approaches actively pursed by the RCS machine designers and engineers for the power upgrades and future machines such as more efficient SC magnets, like 4 T/s ones for the FAIR project in Darmstadt (Germany) built with NbTi SC cables \cite{FAIR} or 12 T/s HTS-based magnet prototype recently tested at FNAL \cite{HTS}; FFAG(fixed field alternating gradient) accelerators \cite{FFAG}; also, the focus is on more efficient power supplies with capacitive energy storage and recovery, and on more economical RF power sources such as 80\% efficient klystrons, magnetrons, and solid-state ones  (compare to current $\sim$ 55\% ) \cite{Yakovlev}. 

The utmost PPP challenge is about how to lower the beam losses while increasing intensity $N_{ppp}$. The tolerable uncontrolled radiation level in accelerator enclosures is typically about $W\sim$1 W/m, so the fractional beam loss must be kept under 
\begin{equation}
\frac{\Delta N_p}{N_p} \leq \frac{W}{P_b} \sim \frac{W}{N_p \gamma},
\label{ploss}
\end{equation}
i.e., the limit should go down with increase of beam intensity, energy and power. On the contrary, repelling forces of the proton beam's own space-charge lead to increase of beam sizes and particle losses at higher beam intensities. In circular accelerators, the empirical {\it space-charge parameter} limit is observed at \cite{Hoffmann}
\begin{equation}
\Delta Q_{SC}= \frac{N_p r_p B_f}{4 \pi \varepsilon_n \beta \gamma^2} \leq 0.3-0.4. 
\label{dQsc}
\end{equation}
Here $r_p$ is classical proton radius, $B_f$ is the bunching factor (ratio of the peak to average bunch current), $\varepsilon_n$ is the normalized beam emittance proportional to the square of the beam size $\sigma^2$. Beyond the limit, the losses grow  unacceptably with increase of the bunch intensity $N_p$ that violates the condition Eq.(\ref{ploss}). 

\begin{figure}[h]
\centering
\includegraphics[width=80mm]{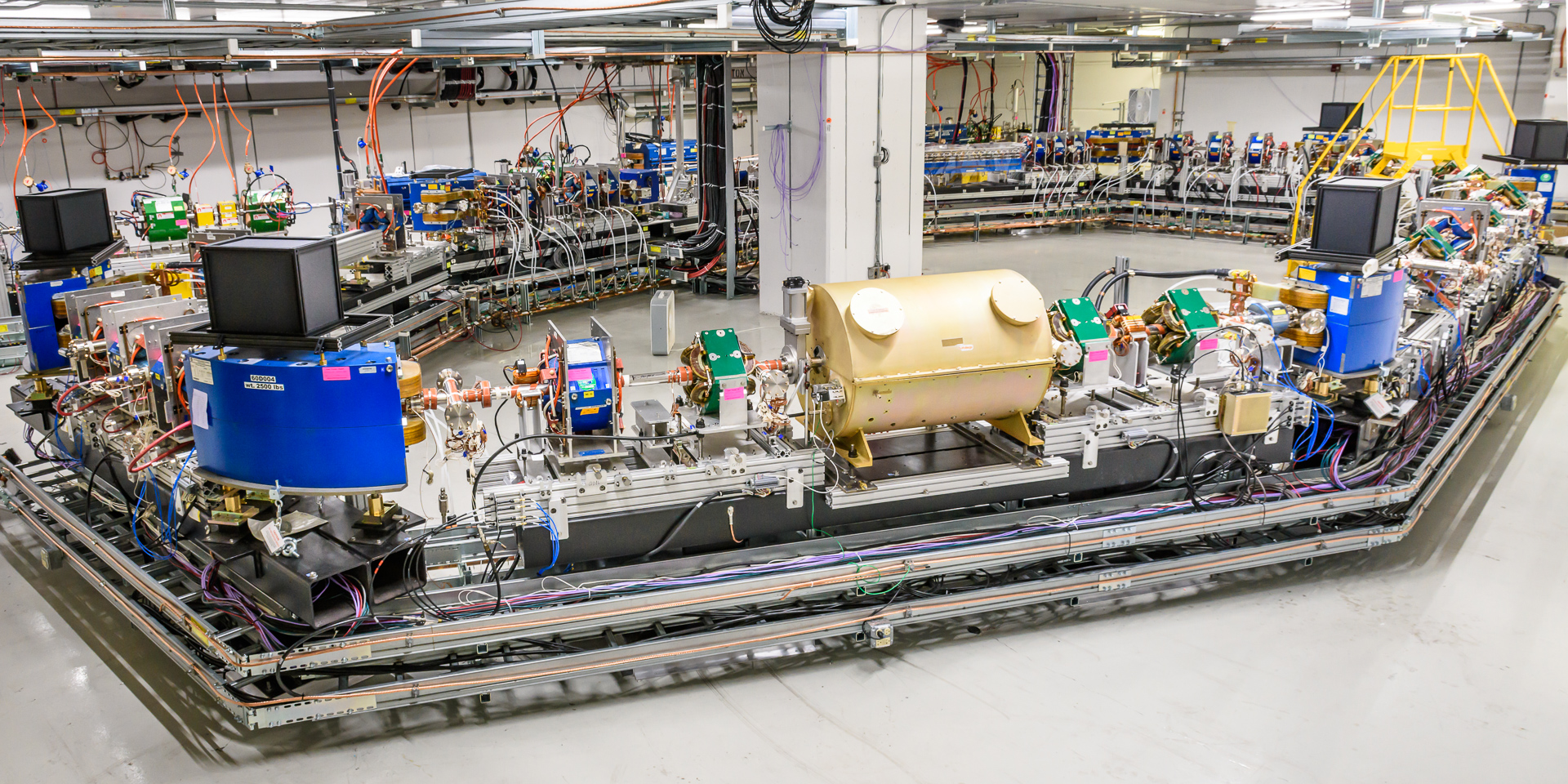}
\caption{Integrable Optics Test Accelerators (IOTA) 40-m circumference ring, capable of operation with 70-300 MeV/c protons and electrons and dedicated for advanced high-intensity beam physics research.} 
\label{IOTAring}
\end{figure}

There are several lines of attack on the problem: i) to increase the injection energy into the RCS - that allows to gain the intensity $N_p \sim \beta \gamma^2$, but requires (often - costly) injection linacs (linacs have an anvantage of much faster average acceleration of about 5-20 MeV/m compared to 0.002-0.01 MeV/m in the rings, thus, protons get through lower energies faster and not being blown up by the space charge forces; one of the limitations of the high intensity linacs is the intrabeam stripping of electrons off H- particles); ii) to employ larger aperture magnets which would allow certain blowup of the transverse rms size $\sigma$ without the beam halo hitting the aperture and resulting in the losses - this method can also be expensive; iii) to flatten the beam current pulse shape by using the 2nd harmonics RF and reduce $B_f$, sometimes by as much as a factor of 2; iv) to use the {\it transverse painting} (shaping) of the beams via charge-exchange injection to linearize the SC forces; v) to design better efficiency collimation system, say, from the current $\sim$60-80\% to 95\% - that will not change the total loss but will result in a lower {\it uncontrolled} irradiation due to the particles which might avoid dedicated collimators or dumps and scattered around the ring; vi) to make the beam focusing lattice perfectly periodic, e.g., like in the Fermilab Booster RCS with periodicity of $P$=24, or in the J-PARC MR with $P$=3 ; vii) to introduce {\it Non-Linear Integrable Optics} elements to  reduce the losses \cite{IO}; viii) to employ electron lenses for the {\it space-charge compensation} \cite{Elens}. The last two approaches are novel and are being tested at the dedicated IOTA ring at Fermilab \cite{IOTA} - see Fig.\ref{IOTAring}. There are other intensity dependent issues in the RCSs for HEP, such as efficient injection of high power beams (that requires stripping electrons off H- particles by foils or lasers), electron cloud effects and coherent beam instabilities. 

In general, RCSs for the neutrino research employ several of these techniques in their quest for higher beam power on target. E.g., the J-PARC accelerators follow the above items ii, iii, iv, vi, while the Fermilab's next complex upgrade, called the Proton Improvement Plan-II (PIP-II) project, is to replace the existing 400 MeV normal-conducting RF pulsed linear accelerator with a new 800 MeV machine based on superconducting RF cavities, capable of CW operation. The PIP-II linac construction is started in the Spring of 2019. The linac will allow an increase of the 120 GeV proton beam power available to the new LBNF beamline to 1.2 MW. In addition, the 8 GeV Booster RCS rate will be increased from 15 to 20 Hz, allowing full Main Injector beam power to be achieved at the lower energy of 60 GeV (vs current 120 GeV), and 80 kW of beam for the 8 GeV neutrino program \cite{PIPII}. PIP-II linac will be part of eventual extension of beam power to LBNF/DUNE to more than 2 MW  (PIP-III - either a SRF linac or an RCS \cite{PIPIII}) and will also provide a flexible platform for long-range development of the Fermilab complex; in particular, provide an upgrade path for a factor of 10 increase in beam power to the Mu2e experiment, and for extension of accelerator capabilities to include flexible high-bandwidth pulse formatting/high beam power operations \cite{MC2018}.

The J-PARC team also explores possibilities for the multi-MW beam power for neutrino experiments, such as construction of a new 8-GeV Booster in addition to their existing 3 GeV RCS to attain 3.2 MW out of the MR and even a new 9 MW power machine with a 9-GeV proton driver consisting of three SRF linacs (1.2 GeV, 3.3 GeV and 6.2 GeV) in the straight sections of the KEKB tunnel which can operate after the conclusion of the Super KEK B-factory project \cite{JPARC2019}.

\begin{table}[h]
\begin{center}
\caption{Main parameters of present and future proton accelerators for high energy neutrino research.}
\begin{tabular}{|l|c|c|c|c|c|}
\hline \textbf{Facility} & \textbf{$E_b$, GeV} & \textbf{$T_{cycle}$, s} &
\textbf{$N_{ppp}$, 10$^{13}$} & \textbf{$P_b$, MW} & \textbf{Year}
\\
\hline FNAL MI & 120 & 1.33 & 5.2 & 0.76 & 2019 \\
 (PIP-II) & 120 & 1.2 & 7.6 & 1.2 & 2026 \\
 (PIP-III) & 120 & 1.2 & 15 & 2.4 & 2030's \\
\hline J-PARC & 30 & 2.48 & 25 & 0.475 & 2019 \\
 & 30  & 1.16 & 43 & 1.3 & 2028 \\
\hline ESS$\nu$SB & 2 & 0.071 & 110 & 5 & ca.2030 \\
\hline ORKA/P & 70 & 7 & 5 & 0.09 & ca.2026 \\
 & 70 & 7 & 25 & 0.45 & ca.2035 \\
\hline $\nu$STORM & 100 & 3.6 & 4 & 0.16 & (tbd) \\
\hline ENUBET & 400 & 5.8 & 2.25 & 0.51 & (tbd) \\
\hline
\end{tabular}
\label{protons}
\end{center}
\end{table}
As mentioned above, issues associated with high power targetry, such as the radiation damage and thermal shock-waves, are critical and depend on pulse structure.  Existing neutrino targets and horns are good to about 0.8 MW beam power, and MW and multi-MW targets are under active development and prototyping. Ongoing R\&D program includes studies of material properties, new forms (foams, fibers), new target designs (e.g., rotating or liquid targets) \cite{Zwaska2018}. In addressing these issues we learn from lower energy but record 1.4 MW beam power machines such as PSI in Switzerland and the SNS in the US.

\section{FUTURE FACILITIES PROPOSALS}
Four proposals of accelerator-based facilities for neutrino physics are presented below. Below, I will reference only the corresponding brief inputs to the 2019 European Particle Physics Strategy Update symposium (EPPSU, May 2019, Granada, Spain) where all of them were presented. More details on each of the proposals can be found therein. Table \ref{protons} summarizes main parameters of these experiments.  

\subsection{Protvino-to-ORKA}
The Protvino accelerator facility is located some 100 km south of Moscow in Russia. It currently consists of a 30 MeV linear  accelerator, and a 1.5 GeV booster synchrotron and the 1.5 km circumference U-70 synchrotron,
which accelerates protons up to 70 GeV.  Typical proton intensity of up to 1.5$\cdot$10$^{13}$ protons per 10 s cycle results in the average beam power up to 15 kW. The proposed upgrade of the injection scheme will make it possible to increase beam intensity to 5$\cdot$10$^{13}$ protons or more to provide some 75-90 kW of the beam power 75 kW with a 7 s cycle. 
With a proper proton beamline and neutrino target to be built, the will allow  to generate and direct a neutrino beam from Protvino towards the KM3NeT/ORCA
detector which is currently under construction in the Mediterranean sea 40 km off shore Toulon, France. The Protvino-to-ORCA experiment, would yield an unparalleled sensitivity to matter effects in the Earth, allowing to determine the neutrino mass ordering with a high level of certainty due to its long baseline of 2595 km after only 5 years of operation after a possible start in 2026 \cite{EPPSU124}. The second phase of the experiment comprizing of a further intensity upgrade of the accelerator complex to 450 kW operation and a significant modification of the ORCA detector would start ca. 2035 and allow for a competitive and complementary measurement of the leptonic CP-violating Dirac phase with a megaton detector.

\begin{figure}[h]
\centering
\includegraphics[width=80mm]{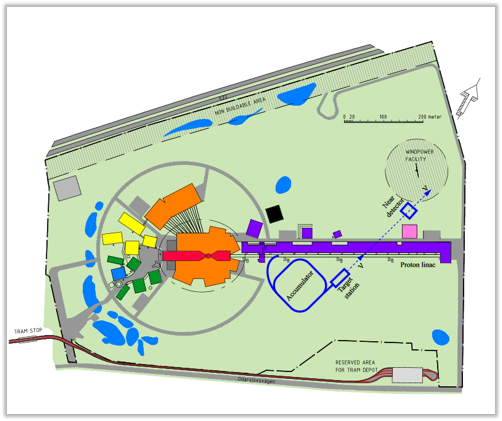}
\caption{Scheme of the ESS Super Beam (ESS$\nu$SB) accelerator complex that includes 2 GeV H- linac, a 400-m circumference accumulator ring, 5 MW target station, and neutrino near detector \cite{Dracos2018}. } 
\label{ESSNUSB}
\end{figure}

\subsection{ESS Neutrino Super Beams (ESS$\nu$SB)}
The 1.83E Euro European Spallation Source (ESS) is currently under construction in Lunbd (Sweden). It will start high power operation in early 2020's and employ a 600 m long SRF 2 GeV proton linac. At the average beam current of 62.5 mA and 4\% duty factor it will deliver 2.8 ms beam pulses with average power of 5 MW on a spallation target. The total site AC power requirement is about 32 MW. The beam power can be raised to 10 MW by increasing the accelerator duty cycle from 4\% to 8\% and the additional 5 MW used to generate a uniquely intense neutrino Super Beam (ESS$\nu$SB) for measurement of leptonic CP violation \cite{EPPSU098}. Tentative schedule calls for the project's CDR in 2021, and TDR 2024 and construction period of 2026-2029. Besides the upgrade of the linac repetition rate from 14 Hz to 28 Hz, it should switch from operation with protons to operation with H- particles. An accumulator ring with 400 m circumference will need to be built to compress to the beam pulse to $\mu$s - see Fig.\ref{ESSNUSB}. Due to very short beam pulse, the required  5MW neutrino target station will be much more challenging than the 5 MW ESS neutron spallation target. One should also expect - and address - very strong space charge effects both in the linac and in the accumulator ring. 

ESS$\nu$SB is expected to be complementary to other proposed Super Beam experiments by the fact that the resulting high intensity $O$(0.3 GeV) neutrino-beam is directed towards the north in the direction  of the Garpenberg mine, 540 km away, which could host the far 1 megaton water Cerenkov detector, at the location of the second neutrino oscillation maximum, making the performance of ESS$\nu$SB for leptonic CP violation precision measurements highly competitive. The total cost of the ESS$\nu$SB is estimated to be 1.3 B Euros. 

\subsection{ENUBET Short Baseline Facility at CERN}
The ENUBET collaboration has proposed a short baseline experiement based on the CERN's SPS proton synchtron to carry out high precision measurements of the neutrino cross sections as function of energy \cite{EPPSU057}. The simplest implementation of this facility can be based on a conventional fast  extraction of the SPS 400 GeV protons, a $\mu$s horn and a $\sim$ 40 m narrow band transfer line. The ultimate average beam power out of the SPS can be as high as about 0.5MW achieved during the CNGS operation. The central energy of secondaries (pions, kaons) will be 8.5 GeV that will result in 0.5-3.5 GeV neutrino’s. 

\subsection{$\nu$STORM}
\begin{figure}[h]
\centering
\includegraphics[width=80mm]{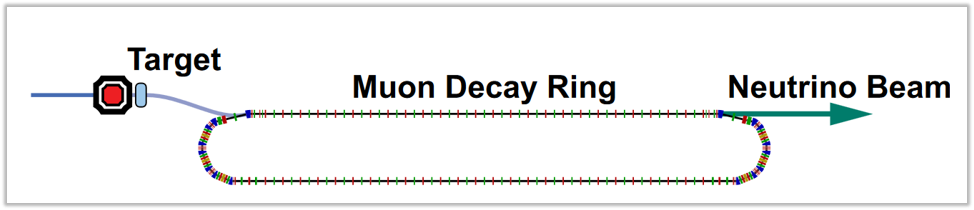}
\caption{Scheme of the $\nu$STORM accelerators : proton beam from the CERN SPS (or other high power proton accelerator) hits the target, the resulting muons are collected and injected into a 585 m racetrack storage ring where they circulate for hundred turns decaying into the very well directed neutrino beam. } 
\label{NUSTORM}
\end{figure}

The Neutrinos from Stored Muons ($\nu$STORM) facility has been designed to
deliver a definitive neutrino-nucleus scattering programme using beams of electron and muon antineutrinos from the decay of muons confined within a storage ring \cite{EPPSU154}. Protons from the CERN's SPS are sent onto a target and converted into muons. The $\mu^{\pm}$ beams with a central momentum of between 1 GeV/c and 6 GeV/c and a momentum spread of 16\% are injected and stored in a racetrack-shaped ring, oriented toward neutrino near-detector some 50 m away and a far-detector 2 km away - see Fig.\ref{NUSTORM}. At $\nu$STORM,
the flavour composition of the beam and the neutrino-energy spectrum
are both precisely known and the storage-ring instrumentation will allow the neutrino flux to be determined to a precision of 1\% or better.

The facility does not call for the record power out of the SPS - it requires only 156kW of 100 GeV protons (4$\cdot$10$^{13}$ protons per pulse in two 10 $\mu$s fast extractions with $T_{cycle}$=3.6 s). The major challenges of the proposal is necessity to have a large diameter (0.5 m) magnets to accept most of the secondary muons and a sophisticated focusing lattice which should assure survival of about 60\% of muons after 100 turns with the 10\%rms beam momentum spread. Cost estimate for such a facility if built at CERN is 160 MCHF. 
 
\section{Summary}
Over the past decade we witnessed impressive progress of the high-energy high- power accelerators for neutrino research - J-PARC facility in Japan has approached 0.5MW of the 30 GeV proton beam power, the Fermilab Main Injector delivers over 0.75MW of 120 GeV protons. The needs of neutrino physics call for the next generation, higher-power, megawatt and multi-MW-class facilities. Some of them are already under construction, such as, e.g., Fermilab’s PIP-II linac upgrade. In general, it is expected that  MW-class and, especially, multi-MW beam facilities will face many challenges. Hence, comprehenisve accelerator R\&D is required and, in many cases -  is started, such as on the cost saving technologies for efficient rapid cycling magnets and RF sources, on the control of space-charge effects, instabilities and beam losses (some counter-measures can be tested at the operational machines and at the IOTA ring at Fermilab), on the multi-MW neutrino targets and horns. There are also new proposals which show significant scientific promise and should be further studied - Protvino/ORKA, ESSvSB, ENUBET, and $\nu$STORM.

\begin{acknowledgments}
I would like to acknowledge very fruitful discussions on the subject of this presentation and thank Yuri Alexahin, Paul Czarapata, Paul Derwent, Jeff Eldred, Steve Holmes, Valeri Lebedev, Sergei Nagaitsev, Bill Pellico, Eric Stern, Cheng-Yan Tan, Alexander Valishev, Bob Zwaska (all – Fermilab), Eric Prebys (UCD), Frank Schmidt (CERN), and David Bruhwiler (RadiaSoft). Some of this material has appeared in other forms written by the author.   

Fermi National Accelerator Laboratory is operated by Fermi Research Alliance, LLC under Contract No. DE-AC02-07CH11359 with the
United States Department of Energy.

\end{acknowledgments}

\bigskip 

\end{document}